\documentstyle[psfrag,graphicx]{llncs}
\newcommand{\be}{\begin{equation}}
\newcommand{\ee}{\end{equation}}

\begin{document}
\pagestyle{empty} \mainmatter

\title{From Tensor Equations to Numerical Code -- 
Computer Algebra Tools for Numerical Relativity}
\titlerunning{Computer Algebra Tools for Numerical Relativity}

\author{Christiane Lechner, Dana Alic \and Sascha Husa}
\authorrunning{Christiane Lechner, Dana Alic \and Sascha Husa}

\institute{Max-Planck-Institut f\"ur Gravitationsphysik,
Albert-Einstein-Institut,\\
 D-14467 Golm, Germany \\
\email{lechner@aei.mpg.de alda@aei.mpg.de shusa@aei.mpg.de}}

\maketitle

\begin{abstract}
     In this paper we present our recent work in developing a computer-algebra  tool for systems of partial differential equations (PDEs), 
termed "Kranc". Our work is motivated by the problem of finding  solutions of the Einstein equations through numerical simulations.
Kranc consists of Mathematica based computer-algebra packages, that facilitate the task of dealing with symbolic tensorial calculations 
and realize the conversion of systems of partial differential evolution equations into parallelized C or Fortran code.



%
%

\end{abstract}

\section{Introduction}

This paper presents our work concerning computer algebra tools to support
the algebraic manipulation and numerical solution of tensorial systems of 
partial differential equations. 
Such systems arise in many areas of physics, our own motivation
is rooted
in the problem of the numerical simulation of compact objects that
emit gravitational radiation, such as black holes or neutron stars.
The underlying theory of gravitation for compact objects is general
relativity.
In general relativity \cite{Wald} the geometry of spacetime is related to the 
matter content via the Einstein equations, $G_{ab} = \kappa T_{ab}$. 
This four-dimensional covariant formulation however is not suited 
for all purposes. E.g.~the task of finding solutions to the equations via 
numerical approximations -- the subject of numerical relativity -- 
rather requires to reformulate the equations such that an 
initial value problem can be solved.

This reformulation of the equations is not unique and
an important problem in numerical relativity is to 
find an evolution system which is well-suited for numerical simulations.
This task involves among others the analysis of non-linear systems of 
partial differential equations that can contain dozens of evolution variables
and to code up expressions which easily can contain up to a thousand terms.
Both to save computational time as well as to reduce possible errors 
it proved very helpful to develop computer algebra tools that facilitate
the analysis of the equations and automatize the generation of numerical code.

Kranc \cite{Kranc}, \cite{ERE2002} is a suite of Mathematica tools and 
packages developed (by the authors together with Ian Hinder) 
to support and automatize the above mentioned steps.
One part -- based on the Mathematica package MathTensor \cite{MathTensor} -- 
aids the analysis and manipulation of tensor equations at the level 
of abstract indices. These tools support e.g.~a decomposition 
of tensor equations, the analysis of the 
principal part of systems of evolution equations or the derivation
of propagation equations for constraint equations. 
The other part provides fully automatized functionality to convert systems of 
evolution equations into parallelized C or Fortran code. 
This second part is described in detail in \cite{Kranc}. 
Our current main application of Kranc-generated codes is a systematic effort to compare
different formulations of the Einstein equations as an evolution system \cite{mexico},
based on standardized testbeds. These tests also serve as our main tool to validate the generated
code.

In the present paper we concentrate on the computer algebra tools 
that facilitate a 3+1 decomposition of four-dimensional 
tensorial equations (including frame based formulations) 
as well as the conversion of tensor equations into equations for individual 
components that can directly be used in a numerical code.
In the following we will also give
a brief overview of the numerical implementation as based on
the Cactus computational toolkit \cite{Cactus}.

The paper is organized as follows: in Sec.~\ref{sec::3+1intro} we give a
brief introduction to the 3+1 decomposition of Einstein's equations, 
and describe the computer algebra tools developed for this purpose in
Sec.~\ref{sec::3+1tools}. Sec.~\ref{sec::frames} describes the extension 
of these facilities to handle frames and in Sec.~\ref{sec::toCode}
the numerical implementation of the resulting evolution equations 
and associated equations is discussed. 

\section{The 3+1 Decomposition of the Einstein
Equations}\label{sec::3+1intro}

\begin{figure}[h]
\begin{center}
\begin{psfrags}
 \psfrag{Sigma}[]{\large $\Sigma_t$}
 \psfrag{Sigmadt}[]{\large $\Sigma_{t+\Delta t}$}
 \psfrag{n}[]{\large $\alpha n^a$}
 \psfrag{beta}[]{\large $\beta^a$}
 \psfrag{t}[]{\large $t^a$}
\includegraphics[width=5in]{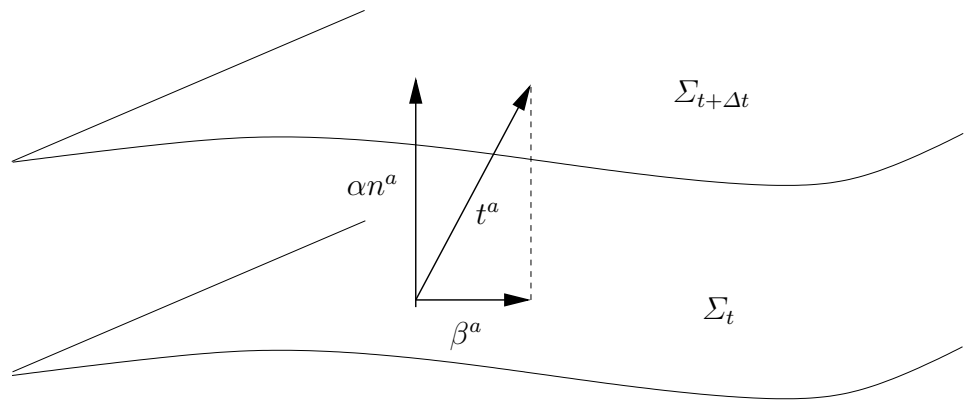}
\end{psfrags}
\caption{Any four dimensional tensor can be decomposed into its parts parallel
and orthogonal to the hypersurface $\Sigma_t$. A prescription of how to move
forward in time (from one hypersurface to the next)
is given by a time-like 
vector field $t^a$ (respectively it's components, 
the lapse function $\alpha = - t^a n_a$ 
and the shift vector $\beta^a = h^a{}_b t^b$)
}\label{fig::3+1}
\end{center}
\end{figure}

One way to formulate the Einstein equations as an initial value problem 
is by foliating
space-time by a family of three-dimensional space-like hypersurfaces $\Sigma_t$
\cite{Wald}.
Denoting the time-like unit normal to the hypersurfaces $\Sigma_t$ by $n^a$, with $n^a n_a = - 1$, the four dimensional metric 
$g_{ab}$ can be written as
\be\label{eq::splitmetric}
g_{ab} = h_{ab} - n_a n_b,
\ee
where $h_{ab}$ is the positive definite metric which $g_{ab}$ induces on $\Sigma_t$.
$(- n_a n^b)$ and $h_a{}^b$ are projection operators in directions orthogonal and tangential to $\Sigma_t$.
Thus any 4-dimensional tensor can be decomposed according to 
\begin{equation}\label{eq::splittensor}
S^{a} = h^{a}{}_{b}S^{b} - n^{a} S,
\end{equation}
with $S = n_b S^b$. Fig.~\ref{fig::3+1} illustrates this decomposition
and shows a time-like vector field $t^a$ which can be interpreted as a
``time-flow'' through space-time.

Splitting the Einstein equations corresponding 
to the above rules yields a set of evolution equations
(PDEs that contain derivatives in the direction of $n^a$) for variables corresponding to tensors tangential to $\Sigma_t$, as well as a set of constraints (PDEs that do not contain time derivatives).
A priory the evolution equations do not have a definite character due to the freedom in specifying the gauge 
(choosing coordinates on the manifold), the freedom to add constraints to the evolution equations as well as choosing different variables.

However a variety of formulations is known for which the evolution system is strongly hyperbolic (in some cases symmetric hyperbolic)
and an evolution system for the constraints can be derived that guarantees that the constraints are satisfied as long as they are 
satisfied initially. In these cases it is sufficient to solve the evolution equations with initial data subject to the constraints 
in order to obtain a solution to Einstein's equations.

Although equivalent on the continuum level, these formulations can behave very differently when solved numerically.
One major open problem in numerical relativity therefore is to determine the formulations and choices of gauge that perform best in 
a given physical situation. This task requires to repeat structurally similar calculations like the 3+1 decomposition mentioned above,
modifying the evolution system, determining characteristics of the evolution system, 
deriving the evolution system of constraints and last but not least implementing the equations into a numerical code for various formulations.

\section{3+1 decomposition using computer algebra}\label{sec::3+1tools}

MathTensor is a Mathematica package for tensor analysis which
for our purpose provides the basic facilities to 
\begin{enumerate}
\item define tensors (with symmetries), 
\item define rules between tensorial expressions 
      (where summation indices are treated with care)
\item the possibility to use several types of indices at the same time 
       (see below)
\item simplify tensor expressions.
\end{enumerate}  
We also use basic geometric concepts like the covariant 
and Lie derivatives, the Riemann tensor etc.  

Kranc supports a 3+1 decomposition (Eqs.(\ref{eq::splitmetric}) and 
(\ref{eq::splittensor})) by providing the following types of functions:
\begin{enumerate}
\item
Functions that define tensors (with symmetries) specified by the user,
labeling them with the attribute ``spatial'' or ``time-like''
(realized by adding the tensor to the appropriate list);
\item Functions that define orthogonality and projection rules for the 
      previously defined spatial tensors, as e.g. for $v_a$: 
      {\tt h[la,ub] v[lb] -> v[la]} and  {\tt n[ua] v[la] -> 0};
\item Functions that implement rules for the basic geometry,
       like decomposing the metric and the covariant derivative of $n^a$;
\item Functions that define rules for the Gauss-Codazzi equations 
      (relating the four dimensional curvature to the three dimensional curvature 
      of the slices $\Sigma_t$); 
\item Functions that convert covariant derivatives in direction of $n^a$ to 
       Lie-derivatives in this direction and split the unit normal $n^a$
      into a time like vector $t^a$ transversal to $\Sigma_t$ and the spatial ``shift'' tangential to $\Sigma_t$.       
\end{enumerate}
Given these facilities the user has to
\begin{enumerate}
\item define her/his spatial variables,
\item define additional rules that decompose four dimensional tensors 
      specific to the system (e.g. the electro-magnetic field in case of the Maxwell equations), 
\item project the four-dimensional equations with 
$n_a n^b$ and $h_a{}^b$,
\item apply the previously defined projection and orthogonality rules. 
\end{enumerate}
These tools have been used to (re-)derive e.g.~the 
Maxwell equations for electro-magnetism, 
the ADM equations for general relativity 
as well as a number of more complicated formulations of the Einstein equations.

\section{Decomposition of a frame-based formulation of the Einstein equations}
\label{sec::frames}

The scope of the present project is the numerical implementation of a frame-based formulation of the Einstein equations.
This system is of special interest to numerical relativity as it has been shown \cite{Friedrich-Nagy} 
that it allows a well posed initial boundary value problem (for a special choice of gauge and hyperbolic reduction).

In this formulation tensors are not expressed by their coordinate components but by their components with respect to an 
orthonormal frame $\{e_I\}$, where $I = 0,1,2,3$. The basic variables of the system are the coefficient functions of the frame $e_{I}{}^{a}$,
the four dimensional connection coefficients $\Gamma_{I}{}^{J}{}_{K}$ defined by
\be\label{connection}
\nabla_{e_I} e_K = \Gamma_{I}{}^{J}{}_{K} e_J
\ee
as well as the Weyl tensor $C_{I J K L}$. The Einstein equations then consist of an equation that relates the Riemann tensor to 
the Weyl tensor, the Bianchi identity  and an equation that ensures that 
the connection is torsion free,
\begin{eqnarray}
R_{IJKL}(\Gamma) & = & C_{IJKL}, \nonumber\\  
\nabla_{I} C_{JKL}{}^{I} &= & 0,  \nonumber\\
\lbrack e_I, e_J \rbrack & = & 
   (\Gamma_{I}{}^{K}{}_{J} - \Gamma_{J}{}^{K}{}_{I}) e_K. 
\end{eqnarray}

Although MathTensor can handle different types of indices, it does not provide any a priori facilities to deal with a frame formalism, 
(e.g.~the connection coefficients (\ref{connection}) are not implemented).  
In order to minimize the effort we decided to rely on Kranc's already 
existing functionality for the projection formalism as far as possible and 
introduce frames only after the equations have been split.

The decomposition of the above system therefore was realized in two steps:
First, we derived equations corresponding to the 3+1 decomposition with 
respect to a time-like unit vector field $e_0$ on a tensorial level. 
Here the existing rules for the projection formalism had to be 
extended to include the case where the time-like vector
$e_0$ is not necessarily hypersurface orthogonal. 
In particular the decomposition of the Riemann tensor is 
slightly more involved in this case.

Second, we extended the time-like unit vector field to an orthonormal basis 
($e_0, e_{i}$) and introduced a second type of index.
One type of index denotes the spatial tetrad indices (i = 1,2,3), the other type denotes the space-time indices (a = 0,1,2,3). 
The conversion from tensor indices to frame indices for spatial objects was realized by rules like $v^a \to v^i e_i{}^a$.
A set of rules introducing the connection coefficients as e.g.~$b^{j}{}_{a} \, e_{k}{}^{c}\, \nabla_{c}\, e_{i}{}^{a} \to \Gamma_k{}^{j}{}_{i}$
completes the transformation to the frame-based formulation.
The evolution equation for the electric part of the Weyl tensor for example
would be given by
\be
e_0{}^a \partial_a E_{ij} 
   + e_l{}^a \varepsilon^{l m}{}_{{\scriptscriptstyle(}i} 
     \partial_{{\scriptscriptstyle |}a{\scriptscriptstyle |}} B_{j{\scriptscriptstyle )} m}  
   - B_{l{\scriptscriptstyle (}i} \varepsilon_{j{\scriptscriptstyle )}}{}^{mn} \Gamma_{m}{}^{l}{}_{n}
   + .... = 0,
\ee
and in MathTensor syntax
\begin{verbatim}
e0[u1]*OD[El[ali, alj], l1] 
   - (b[au1, u1]*Epsilon[al1, ali, au3] OD[B[alj, al3], l1])/2  
   - (b[au1, u1]*Epsilon[al1, alj, au3]*OD[B[ali, al3], l1])/2 
   - B[al2, ali]*Epsilon[al3, alj, au4]*gamma[au3, al4, au2])/2
   + .... == 0,
\end{verbatim}
where in the Mathematica code the variables $E$ and $\Gamma$ have been
renamed to {\tt El} and {\tt gamma}.

The system thus obtained has to be supplemented by a choice of gauge
and it may be necessary to add constraints to the evolution equations in 
order to obtain a symmetric hyperbolic system of equations.
 
In particular, for the formulation in
Ref.~\cite{Friedrich-Nagy} the gauge is adapted to a time-like boundary
and to a foliation of the domain of computation 
by time-like hypersurfaces. According to that a further ``2+1 split'' of the 
``3+1 equations'' has to be performed. Certain constraints are added to the 
evolution equations. The resulting evolution system is symmetric hyperbolic
and maximally dissipative boundary conditions can be specified at the 
time-like boundaries. Furthermore the constraints are 
satisfied by virtue of the evolution system if they are satisfied on the
initial hypersurface, irrespective of the boundary conditions for the 
main evolution system.

\section{Converting tensor equations to C or Fortran code}\label{sec::toCode}

The equations in abstract index notation given in MathTensor 
syntax are converted to equations that 
can be used in a numerical code by the following steps: 
\begin{enumerate}
\item Summation indices of both types are expanded using the MathTensor 
      command {\tt MakeSumRange}. 
\item Kranc provides functions that split tensorial expressions into a list 
      of independent components, e.g.~a symmetric 2-tensor $E_{i j}$ would 
      be split into a list of components 
      $\{E_{11}, E_{21}, E_{22}, E_{31}, E_{32}, E_{33}\}$.
\item Names are assigned to the components of tensors as they 
      should appear in a numerical code as e.g.~$E_{11} \to$ {\tt E11}. 
\item Partial derivatives are assigned standard names such as 
      $\partial_1 B_{12} \to$ {\tt D1[B12]}.
\end{enumerate}

The resulting equations can be converted automatically into numerical code, 
which is integrated into the Cactus computational toolkit \cite{Cactus}.
Cactus is an open source problem solving environment designed for 
scientists and engineers.
Cactus mainly targets the issues of parallelization, modularization and 
portability. The name Cactus derives from the design of a central core 
(or "flesh") which connects to application modules (or "thorns") through an 
extensible interface. Thorns can implement problem specific code on any 
level from concrete physics applications to low-level infrastructure.

A set of thorns comprising the ``Cactus computational toolkit'' provide
infrastructure such as parallel I/O, data distribution and checkpointing. 
Current data distribution implementations are based on the 
principles of domain decomposition and message passing 
using the MPI standard \cite{MPI}.
Using Cactus yields an open and reasonably {\em documented} infrastructure,
parallelization, a variety of I/O methods and allows easy interfacing
with a growing community writing numerical relativity Cactus applications.

Cactus does not have the structure of a library which provides a set of
functions that can be called by the user application. 
Instead, all Cactus thorns are compiled into libraries,
and management tasks such as code execution and allocation of distributed
memory are handled by Cactus and steered via configuration files. 
The end user supplies a ``parameter file'' which specifies all 
user-controllable aspects of the run.

The basic module structure within Cactus is called a ``thorn''. All
user-supplied code is organized into thorns, 
which communicate with each other via calls to the flesh API
(application programmer interface) or APIs of other thorns. 
The integration of a thorn into the flesh or with other thorns is specified
in configuration files which are parsed at compile time.

Kranc provides fully automatized functionality 
to generate complete Cactus thorns. This is described in detail in 
\cite{Kranc}. In the following we only briefly summarize 
the basic settings of the numerical implementation. 
The setup for an evolution system (first order in time) together with 
associated equations (like e.g.~the constraints) as 
described above would be the following:
\begin{enumerate}
\item The Kranc function {\tt CreateBaseThorn} generates a Cactus 
      ``Base thorn'' to declare grid functions and parameters. 
\item Time integration is based on the MoL
      \cite{CactusMoL} method of lines thorn within Cactus, developed by 
      I.~Hawke.
      This code provides a parallel ODE integrator, implementing 
      generic Runge-Kutta and iterative Crank-Nicolson methods.
      The Kranc function {\tt CreateMoLThorn} is used to generate
      a thorn that calculates the right hand sides of the evolution equations.
\item Quantities that enter the evolution equations but are not 
      evolved variables (like e.g.~gauge source functions, 
      which are prescribed as functions of the coordinates) 
      are set at every time step by means of a ``Setter thorn''.
      A Setter thorn that is only called initially can be used to 
      specify initial data. The Kranc function {\tt CreateSetterThorn}
      generates such thorns.
\item The Kranc function {\tt CreateEvaluatorThorn} 
      can be used to generate thorns to evaluate the constraints 
      at time steps specified at run time.  
\item For applications in numerical relativity a set of minimal initial data
      can be computed with the help of already existing Cactus thorns. 
      In this case a ``Translator thorn'' generated by 
      {\tt CreateTranslatorThorn} transforms these data to
      the evolution variables used in the present formulation.  
\item Currently Kranc supports only simple boundary conditions 
      (as e.g.~periodic boundary conditions). 
      An extension to maximally dissipative boundary conditions 
      is planned for the near future. 
\end{enumerate}

We conclude this section by giving an (abbreviated)
example of code, generated by the function {\tt CreateMolThorn},
which assigns the right hand sides for the MoL evolution 
for the Weyl system (the gauge has been simplified such that the
vector $e_0$ is hypersurface orthogonal).

\begin{verbatim}
// several header files are included and some macros get defined
/* Define macros used in calculations */  
               // C++ style comments do not appear in original code
#define INITVALUE  (42)

void KrancFNMoL_CalcRHS(CCTK_ARGUMENTS)
{
 DECLARE_CCTK_ARGUMENTS
 DECLARE_CCTK_PARAMETERS

 /* Declare the variables used for looping over grid points */
 int i = INITVALUE;  // same for j, k, index, istart, jstart, 
                     // kstart, iend, jend, kend 

 /* Declare finite differencing variables */
 CCTK_REAL dx = INITVALUE;  // same for dy, dz, dxi, dyi, dzi, 
                            // hdxi, hdyi, hdzi

 /* Declare shorthands */ // none in this example

 /* Declare local copies of grid functions */
 CCTK_REAL B11L = INITVALUE;   // same for B21L, B11rhsL, D1B11, etc.

 /* Initialize finite differencing variables */
 dx = CCTK_DELTA_SPACE(0);
 dxi = 1 / dx;
 hdxi = 0.5 * dxi;  // analogously for x and y directions

 /* Set up variables used in the grid loop 
 /* with stencils suitable for finite differencing */
 istart = cctk_nghostzones[0];
 iend   = cctk_lsh[0] - cctk_nghostzones[0]; // analogously for jstart, 
                                             // kstart, jend, kend

 /* Loop over the grid points */
 for (k = kstart; k < kend; k++)
 {
  for (j = jstart; j < jend; j++)
  {
   for (i = istart; i < iend; i++)
   {
    index = CCTK_GFINDEX3D(cctkGH,i,j,k); // a Cactus macro to 
                                          // compute the 3D array index

    /* Assign local copies of grid functions */
    B11L = B11[index]; // analogous for B21L etc.

    /* Precompute derivatives */
    D1B11 = D1gf(B11,i,j,k);  // analogous for D1B21 etc.

    /* Calculate grid functions */
    B11rhsL  =  2*B11L*chi11L + 2*B31L*chi13L + B21L*chi21L - 
       B22L*chi22L - B32L*chi23L + B31L*chi31L - B32L*chi32L + 
       B11L*chi33L + B22L*chi33L - El21L*gamma131L - 
       El32L*gamma221L + 2*El11L*gamma231L + El22L*gamma231L - 
       El11L*gamma321L + El22L*gamma321L + El32L*gamma331L + 
       El31L*gamma332L - 2*B11L*trKL - e31L*D1(El21,i,j,k) + 
       e21L*D1(El31,i,j,k) - e32L*D2(El21,i,j,k) + 
       e22L*D2(El31,i,j,k) - e33L*D3(El21,i,j,k);  // etc.

    /* Copy local copies back to grid functions */
    B11rhs[index] = B11rhsL;  //etc.
   }
  }
 }
}
\end{verbatim}


\begin{thebibliography}{9}
\bibitem{CactusMoL} The MoL thorn is part of the standard Cactus distribution
         \cite{Cactus}.
\end{thebibliography}
\end{document}